\documentclass[usenatbib]{mnras}

\usepackage[T1]{fontenc}
\usepackage{aecompl}
\usepackage{amsmath,amsfonts,amssymb}
\usepackage[pdftex]{graphicx}
\usepackage{aas_macros}
\usepackage[usenames]{xcolor}
\usepackage{caption,color}
\usepackage{gensymb}
\usepackage{subfig}
\usepackage{natbib}

\title[Planet-like spirals caused by moving shadows in TDs]{Planetary-like spirals caused by moving shadows in transition discs}
\author[Montesinos \& Cuello]{Mat\'ias Montesinos$^{1,2}$\thanks{Email: mmontesinos@gmail.com } \& Nicol\'as Cuello$^{1,2}$\\ 
 $^{1}$ Instituto de Astrof\'isica, Pontificia Universidad Cat\'olica de Chile, Santiago, Chile\\
 $^{2}$ Millennium Nucleus ``Protoplanetary discs'', Chile}

\begin{document}

\defcitealias{M+2016}{M+16}

\date{Accepted ... Received ...}

\pagerange{\pageref{firstpage}--\pageref{lastpage}} \pubyear{2017}

\maketitle

\label{firstpage}

\begin{abstract}
Shadows and spirals seem to be common features of transition discs. Among the spiral-triggering mechanisms proposed, only one establishes a causal link between shadows and spirals so far. In fact, provided the presence of shadows in the disc, the combined effect of temperature gradient and differential disc rotation, creates strong azimuthal pressure gradients. After several thousand years, grand-design spirals develop in the gas phase. Previous works have only considered static shadows caused by an inclined inner disc. However, in some cases the inner regions of circumbinary discs can break and precess. Thus, it is more realistic to consider moving shadow patterns in the disc. In this configuration, the intersection between the inner and the outer discs defines the line of nodes at which the shadows are cast. Here, we consider moving shadows and study the resulting circumbinary disc structure. We find that only static and prograde shadows trigger spirals, in contrast to retrograde ones. Interestingly, if a region of the disc corotates with the shadow, a planet-like signature develops at the co-rotation position. The resulting spirals resemble those caused by a planet embedded in the disc, with similar pitch angles.
\end{abstract}



\section{Introduction}

	The most recent spatially-resolved observations of protoplanetary discs have revealed intriguing structures such as spirals, vortices, shadows, and gaps. Observations of spirals and shadows in discs \citep{Garufi+2013, Benisty+2015, Stolker+2016, Perez+2016, Benisty+2017} show a broad range of morphologies, with apparently moving features in some cases. For instance, in HD 135344B, \cite{Stolker+2017} report temporal variations in the azimuthal brightness distribution of all epochs. The dark narrow lanes in the disc are interpreted as shadows exhibiting variations up to 10\% in the JHK bands.
	
	Shortly after these discoveries, many physical processes and scenarios were proposed in order to reproduce these exotic features in discs. In particular, the most common spiral-triggering mechanisms considered are: planet torques \citep{Dong+2015}, self-gravity effects \citep[and references therein]{Kratter&Lodato2016} and stellar encounters \cite{Pfalzner2003}. However, local thermodynamic effects due to shadowing patterns in the disc can also cause spirals in the gas distribution as shown by \cite{M+2016}. 	In this context, the case of HD142527 is of particular interest. In this well-studied circumbinary disc, spirals and shadows have been observed by \cite{Christiaens+2014} and \cite{Avenhaus+2014} respectively. In order to explain the two diametrically opposed dips in scattered light observations, \cite{Marino+2015} proposed a scenario where the shadows are cast by an inclined optically thick inner disc able to  block a large fraction of the stellar irradiation 
in two azimuthal sectors of the outer disc. More recently, \cite{F+2017} showed that for inclined precessing inner discs, asymmetric shadow patterns can appear in scattered light observations (cf. their figure 9). This happens for modestly inclined inner discs compared to the outer one ($\approx30\degree$), and it is caused by the interaction with a stellar companion. Surprisingly, in this configuration, the position angle of the line connecting the shadows cannot be directly related to the position angle of the inner disc as shown by \cite{Min+2017}. However, provided that the inclination angle of the outer disc is known, this peculiar effect allows to constrain the inclination and the position angle of the inner disc.
	
	In addition to these observation-oriented works, \cite{Kama+2016} studied the geometric effect of a rotating light-obstructing clump in the disc, which results in spiral-shaped shadow patterns. This mechanism, dubbed as ``azimuthal lag'', is solely due to photon traveling effects since no thermodynamics are considered whatsoever. For transition discs with cavities of several tens of au, this produces observable effects for clumps on Keplerian orbits typically below 0.1~au (cf. their figure~1). In this work, we assume that the shadow-casting material is located further away from the central star, hence we neglect any azimuthal lag.

	Here, we aim to answer the following questions: how does the shadow rotation affect the triggering and the structure of the resulting spirals? Is there a difference between prograde and retrograde shadows? What are their characteristic observational signatures? In Section~\ref{sec:theory} we present the equations that describe the moving shadows, and explain how they can be modeled through hydrodynamic simulations. In Section~\ref{sec:results} we present our results, which are then discussed in Section~\ref{sec:discussion}. Finally, in Section~\ref{sec:conclusions}, we draw our conclusions in the context of planet formation.

\section{Moving shadows simulations}
\label{sec:theory}

	In Section~\ref{sec:precession}, we detail the geometry of the binary system considered and the method by which we estimate the precession period of the inner disc, noted $T_{\rm p}$. In Section~\ref{sec:hydro} we describe the disc geometry and the numerical setup of our hydro-simulations.
	
\subsection{Precession period of the inner disc}
\label{sec:precession}

\begin{figure}
\begin{center}
\includegraphics[width=8cm, height=5cm]{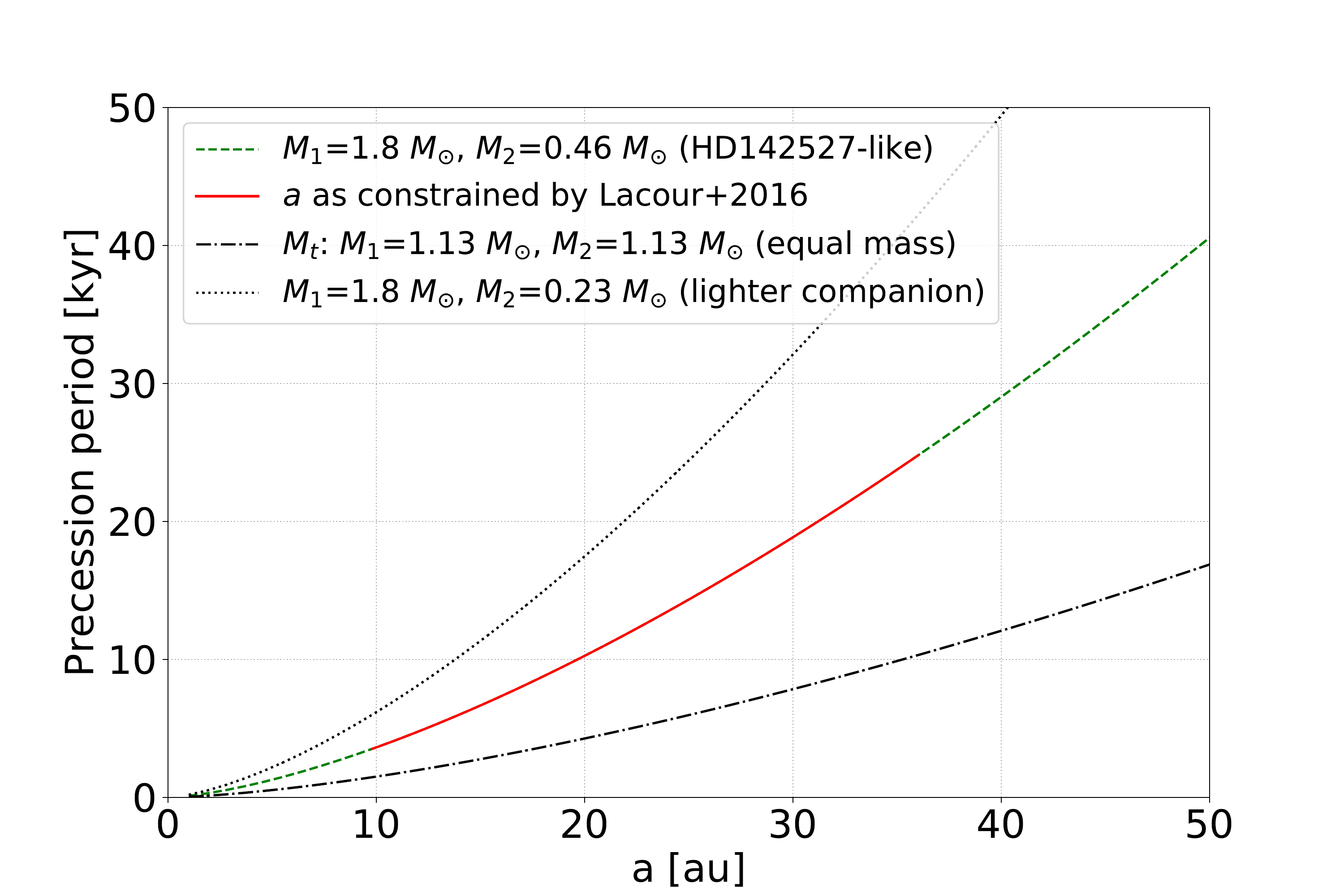}

\caption{Precession periods as a function of the binary semi-major axis for different stellar masses.}
\label{fig:Tp}
\end{center}
\end{figure}

	Here we consider the evolution of a circumbinary disc that has broken due to the interaction with an inner stellar companion, presumably on an inclined orbit. In the disc-breaking scenario \citep{Papaloizou&Pringle1983, Dogan+2015, F+2017}, there are two discs with different inclinations and the (light-obstructing) inner disc is subject to precession. Under our assumptions, this precession translates into a rotation of the line of nodes\footnote{defined as the intersection of the two disc planes.} at the same frequency. Hence, the shadows' ($T_{\rm s}$) and the inner disc' precession periods are equal: $T_{\rm s} = T_{\rm p}$. The inner disc precession depends on both the disc properties and the orbital parameters of the binary \citep{Nixon+2013, F+2017}, noted with the subscripts ``d'' and ``b'' respectively. Interestingly, $T_{\rm p}$ can be expressed in a simple manner by introducing the two following factors:
\begin{eqnarray}
	f_{\rm b} &=& \frac{4}{3 \mu_1 \mu_2 \cos \tilde \beta_{\rm in}} \label{eq:fb} \\
	f_{\rm d} &=& \left( \frac{p+1}{5/2-p} \right) \left( \frac{r_{\rm in}}{a} \right)^{7/2}
				\frac{\left( r_{\rm break} / r_{\rm in} \right)^{5/2 - p} - 1}
				{1 - \left( r_{\rm break} / r_{\rm in} \right)^{-(p+1)} } \label{eq:fd}
\end{eqnarray}				
where $-p$ is the exponent of the surface density power-law (later defined as -1), $a$ the binary semi-major axis and $\mu_i = M_i/ (M_1 + M_2)$ with $i=\{1,2\}$. The angle $\tilde \beta_{\rm in} = 50^{\circ}$ is defined as the average angle between the inner disc and the binary plane during the first precession period \citep{F+2017}. $r_{\rm in}$ and $r_{\rm break}$ are respectively the inner radius of the circumbinary disc \citep{ArtymowiczLubow1994}, and the radius at which the disc breaks\footnote{this happens when the torque exerted by the companion is larger than the internal torques generated by pressure forces.}. For an inviscid disc, \cite{Nixon+2013} report that $r_{\rm break} \approx 2.8 \, a$. Then, according to equation~4 in \cite{F+2017}, the inner disc precession $T_{\rm p}$ reads:
\begin{equation}\label{eq:Tp}
	T_{\rm p} = f_{\rm b} \, f_{\rm d} \, T_{\rm b}
\end{equation}
where $T_{\rm b}$ is the binary period. The only relevant binary parameters in Eq.~\ref{eq:Tp} are the stellar masses ($M_i$) and the semi-major axis $a$. 
Interestingly, under \cite{F+2017} assumptions, $f_{\rm b}$ depends on the binary inclination trough the $\tilde \beta_{\rm in} $ parameter, and independently 
of the eccentricity $e$. In fact, here we assume for simplicity that $r_{\rm break}$ and $r_{\rm in}$ do not depend on these quantities, and that they can both be written as a multiples of $a$: $r_{\rm in} = 1.5 \, a$ \citep{ArtymowiczLubow1994} and $r_{\rm break} = 3 \, a$ \citep{F+2017}. Given that the binary period is given by the Kepler's third law:
\begin{equation}\label{eq:Tb}
	T_{\rm b}^2 = \frac{4 \pi^2}{G \, (M_1 + M_2)} \, a^3 \,\, ,
\end{equation}
we find that $T_{\rm p} \propto  (M_1 + M_2)^{3/2} / M_1 \, M_2 $ and $T_{\rm p} \propto a^{3/2}$. In Figure~\ref{fig:Tp}, we plot $T_{\rm p}$ as a function of $a$ for different pairs of binary masses. We set $M_1 = 1.8 \, M_\odot$ (in agreement with \citealt{Casassus+2015}) and the mass ratio to 0.25, i.e. $M_2=0.46$. We note that, for a fixed value of $M_1$, an increase (decrease) in $M_2$ translates into an decrease (increase) in $T_{\rm p}$. In addition, we also see that, if the binary masses 
are not well constrained (ranging between $0.2 - 2 M_\odot$), for a fixed value of $a$ there is a wide range of values allowed for $T_{\rm p}$, comprised between 1 and 50 kyr approximately. These values justify the shadows' periods used in our simulations.

\subsection{Hydrodynamical simulations with shadows} 
\label{sec:hydro}

	For this study, we use our modified version of {\sc fargo-adsg} \citep{Baruteau+Masset2008}, which models an irradiated disc in the presence of shadows \citep{M+2016}. We consider a 2D self-gravitating gaseous disc orbiting an HD142527-like binary system, i.e. $M_{\rm 1} = 1.8 \, M_\odot$ and $M_{\rm 2} = 0.46 \, M_\odot$, with $a$ ranging from 10 to 35 au approximately. For simplicity, we assume a point-like central potential with mass $M_\star = 2.26$ (instead of a binary one) and a luminosity $L_\star = 23 \, L_\odot$ \citep{Avenhaus+2014}.

The computational domain in physical units extends from $r_{\rm in} = 70$ to $r_{\rm out} = 1000$ au over $n_r = 512$ logarithmically spaced radial cells. The grid samples $2 \pi$ in azimuth with $n_\theta = 1024$ equally spaced sectors.  The initialization of the disc is performed assuming radiative equilibrium, i.e. the energy received from the star equates the energy loss irradiated through the disc's surface, implying an initial flared disc with aspect ratio $h = 0.03 ~ (r/\rm{au})^{0.14}$, needed to reach a quasi-steady state in short time-scales (less than $5000 \rm ~  yrs$). The density is initialized with a low-density inner cavity up to 90 au, and a power-law profile given by:
\begin{equation}
\Sigma(r) = \begin{cases} 10^{-3}   {\rm g \, cm^{-2}},   \,\, & \rm{if }  \,\, 70 \, \rm{au}  \leq  r < 90 \,  \rm{au} \\ 0.62 \, {\rm g \, cm^{-2}} \, \times \left( \frac{r}{\rm 90 \, au} \right) ^{-1} \,\, & \rm{if } \,\,  r \geq 90 \, \rm{au.} \end{cases}
\end{equation}
By doing so, the total disc mass is equal to 0.017 $M_\odot$. We model an $\alpha$-disc \citep{Shakura+Sunyaev1973}, with the $\alpha$ parameter set to $10^{-4}$,
and only heated by stellar irradiation. For simplicity, we take a constant opacity $\kappa = 1 ~ \rm cm^2 g^{-1}$, which fits well for low temperature discs (e.g.,  \citealp{Bell+Lin1994}, \citealp{Bitsch2013}).  

We implement the shadows in the disc in the same fashion as in \cite{M+2016}. We  assume the presence of an inner disc inside the large cavity\footnote{not included in the simulation}, with a certain inclination $i$ with respect to the circumbinary disc. In the previous work by \cite{M+2016}, it was assumed that the circumbinary and the inner discs were face-on and edge-on respectively, i.e. $i=90\degree$, with static cast shadows. However, \cite{F+2017} recently showed that, after the disc breaking phase, the angle $i$ could vary between $20\degree$ and $110\degree$ over time-scales of the order of several hundred binary periods. For simplicity we will assume here that $i \approx 70\degree$, and that inclination is high enough to produce two diametrically-opposed shadows of a certain angular width (cf. their figure~8 for example). This means that there are two moving shadows in the disc, for which the stellar radiation is practically null.

The shadows rotate as a rigid body in either a prograde or a retrograde motion with respect to the sense of the rotation of the gas. Initially, they are set at angles $0\degree$ (right side) and $180\degree$ (left side), with an angular width of approximately $10\degree$. Once the simulation starts, the shadows begin to rotate at fixed angular frequency $\omega_0$. Their azimuthal speed is given by $v_{\rm s}(r) = \omega_0 r$, where $\omega_0 = 2\pi / T_{\rm p}$ (cf. Eq.~\ref{eq:Tp}). Further details about the shadows implementation and the non-stationary energy equation can be found in \cite{M+2016}.

\section{Results}
\label{sec:results}

\begin{figure*}
{\includegraphics[width=1\linewidth]{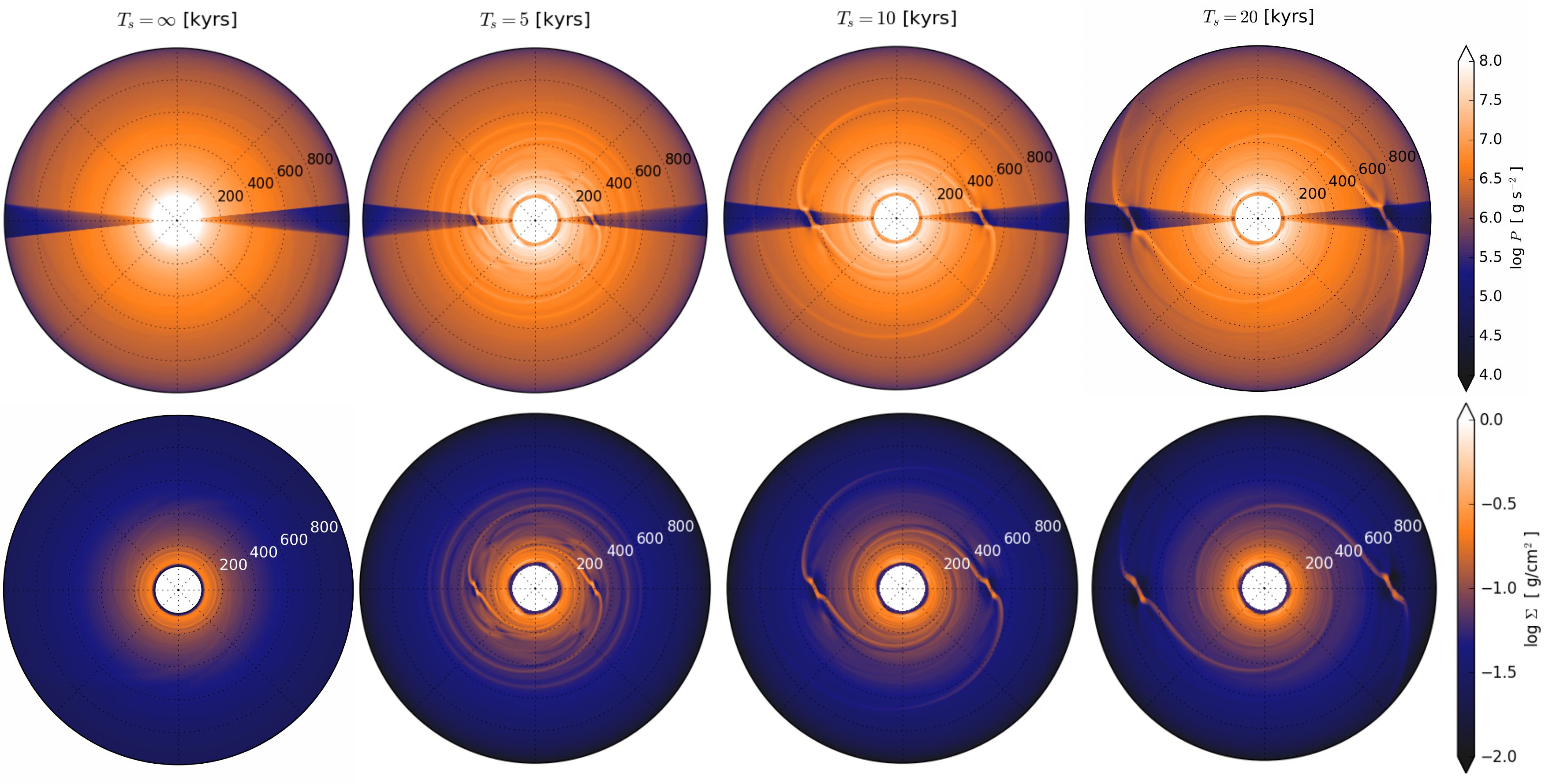}}
\caption{Snapshots taken after 20 kyrs of disc evolution, of the pressure (top) and  surface density field (bottom) for models with shadows rotating at different 
periods: $T_s = \infty$ (static), $5$, $10$ and $20$ kyr. }
\label{fig:dens_press}
\end{figure*}

We study the triggering of spirals in discs with shadows rotating in retrograde, static, and prograde motions. In Table~\ref{tab:simus} we report the set of hydrodynamical simulations for which we observe spirals. Interestingly, we see that retrograde shadows do not lead to spiral formation in the disc\footnote{unless the rotation period is extremely low, i.e. almost static.}. In Figure \ref{fig:dens_press}, we plot the integrated pressure (top) and the surface density (bottom) fields for the static and prograde shadows, after $20$ kyr of disc evolution. At this evolutionary stage, the shadows are projected at angles $0\degree$ and $180\degree$. It is worth mentioning that spirals develop very early (after $\sim $5 kyrs),  with a morphology that sustains in time. The observed asymmetries in Figure \ref{fig:dens_press} are solely due to shadowing effects. In fact, in these low-mass disc models, the disc's self-gravity is not strong enough to trigger fragmentation or any kind of structures. In addition, in simulations without shadows, the density field remains practically unchanged as reported by \cite{M+2016}. 

In Figure \ref{fig:dens_press}, we notice that all the prograde simulations exhibit a \textit{peculiar perturbation} in both the density and  the pressure fields, which are rather similar to the spiral wakes created by a planet. The radial distance of this local perturbation increases as we increase the period of the rotating shadows $T_{\rm s}$. For the model with $T_{\rm s} = \infty$ (fixed shadows), no perturbation is observed, just a weak spiral feature as already shown by \cite{M+2016}\footnote{the present model is 10 times less massive than the model in \cite{M+2016}, explaining the fainter spiral.}. In Table~\ref{tab:simus} we report the radial co-rotating position of the locally induced perturbation ($R_{\rm co}$), for each value of $T_{\rm s}$. The relation between both quantities is discussed in Section~\ref{sec:discussion}. In order to compare the shadows-triggered spirals to planetary ones, we also ran a simulation with a planet of 1~Jupiter mass on a circular orbit at a radius equal to 464 au. This corresponds to the co-rotating location in the prograde shadow simulation with $T_{\rm s} = 10$ kyr. We fit the spiral waves in the density field (Fig.~\ref{fig:dens_press})
with a generalized Archimedean spiral of equation $r(\theta)=A_0 + A_1 \, \theta^n$, where $\theta$ is the azimuth and $\{A_0, A_1, n\}$ are the fitting parameters. We perform this analysis for the outer and the inner (if present) spirals. On the one hand, we observe that all the outer spirals have similar pitch angles ($|\phi| \sim 16\degree$), but the simulation with $T_{\rm s}=20$ kyrs ($|\phi| \sim 20\degree$). On the other hand, we find that the inner spirals for the moving shadows ($|\phi| \sim 23\degree$) are bit larger than the planetary ones ($|\phi| \sim 14\degree$). Again, we see that the pitch angle increases with $T_{\rm s}$ (see Table \ref{tab:simus}). 

\begin{figure}
\begin{center}
\includegraphics[width=0.4\textwidth]{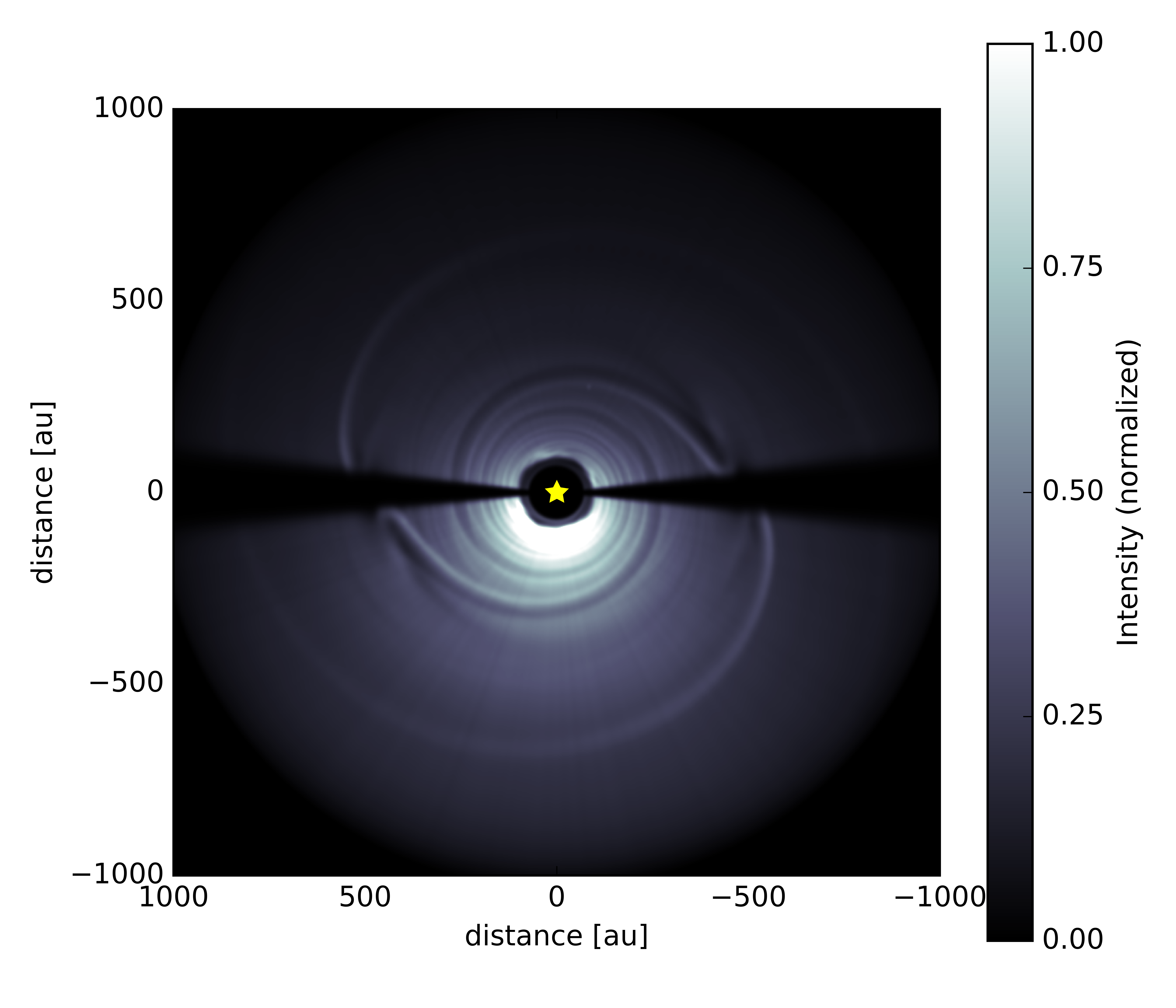}
\caption{Radiative transfer calculation at 1 $\mu$m for the simulation of prograde shadows with $T_{\rm s}=10$ kyr, shown in the third column of Fig.~\ref{fig:dens_press}. We assume a dust-to-gas ratio equal to 0.01.}
\label{fig:obs}
\end{center}
\end{figure}

In Fig.~\ref{fig:obs} we show a synthetic image of the emission of the disc at a wavelength equal to 1~$\mu$m, where the intensity has been normalized and scaled by $r^2$. This calculation is done by means of the {\sc radmc-3d} code \citep{radmc3d}, modifying it to mimic the shadows by blocking the photons sent from the star in the shadowed regions.
It is based on the density field for the simulation with prograde shadows with $T_{\rm s} = 10$ kyr,  assuming a dust-to-gas ratio of 0.01.

\begin{table*}
\begin{center}
\tiny 
\caption{Set of simulations after 20 kyrs  of disc evolution with shadows, for different shadow's rotating  frequencies.}
\label{tab:simus}
\begin{tabular}{|c|c|c|c| |c|c|c|c|c|c|c|c|c|}
\hline
	$T_{\rm sh}$ (kyr) & $R_{\rm co}$ (au) & $A_{\rm in}$ & $B_{\rm in}$ & $n_{\rm in}$ & $\phi_{\rm in}(\degree)$ & | &
	$A_{\rm out}$ & $B_{\rm out}$ & $n_{\rm out}$  & $\phi_{\rm out}(\degree)$\\
\hline
     $\infty$ & n/a & n/a & n/a & n/a & n/a & | & 603 & 194 & 1.24 & -15.52 \\
     5 & 292 & 285 & -93 & 0.35 & -23.22 & | & 307 & 88 & 0.57 & -15.15 \\
     10 & 464 & 431 & -136 & 0.47 & -22.09 & | & 484 & 136 & 0.46 & -16.16 \\
     Jupiter mass planet  & 464 & 389 & -103 & 0.58 & -14.39 & |  & 498 & 157 & 0.72 & -17.21 \\ 
     20 & 717 & 635 & -220 & 0.66 & -28.06 & | & 764 & 178 & 0.55 & -19.95 \\	
\hline
\end{tabular}
\end{center}
\end{table*}

\section{Discussion}
\label{sec:discussion}

\begin{figure*}%
    \centering
    \subfloat{{\includegraphics[width=6cm, height=4.3cm]{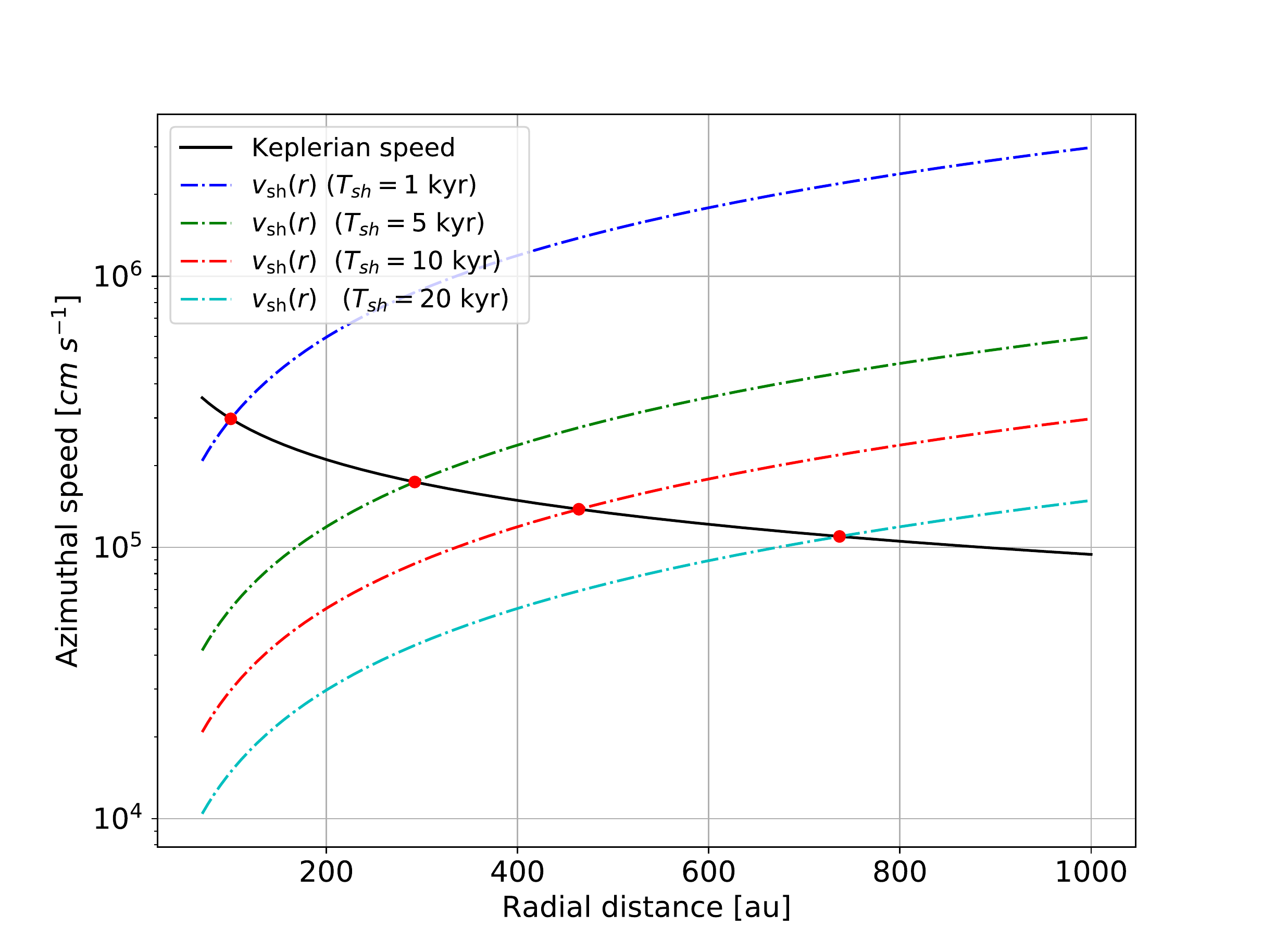} }}%
    \qquad
    \subfloat{{\includegraphics[width=6cm, height=4.3cm]{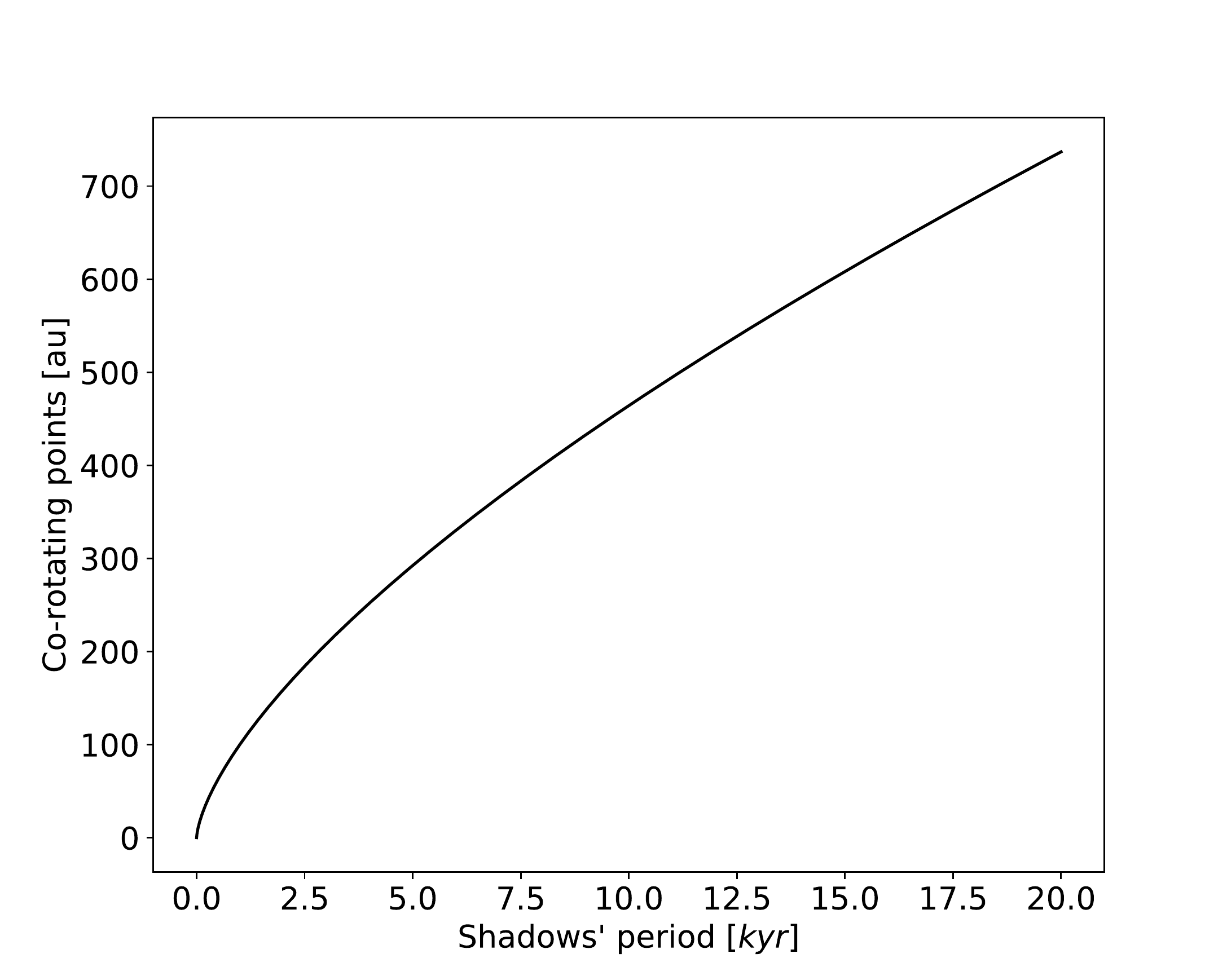} }}%
    \caption{\textit{Left:} Shadow rotation speed ($v_{\rm s} = 2\pi/T_{\rm s} ~ r$) for different periods $T_{\rm s}$ (colored dotted lines) along with the azimuthal velocity (solid curve). The intersections (red dots) give the co-rotating locations, in remarkable agreement with the location of the perturbations in our simulations. \textit{Right:} Co-rotating location as a function of the shadow's period for prograde rotations.}%
    \label{fig:resonance}%
\end{figure*}

Moving shadows dramatically changes the disc structure and evolution. For retrograde rotations, we do not observe any spirals, despite the fact that there are two shadows cast in the disc. Because the shadows are always confronting the gas motion, there is not enough time for it to cool down in order to create a gradient pressure to act as a symmetry breaking. This could be different if a more efficient cooling mechanism is taken into account (e.g., $\beta$ cooling prescription, \citealt{Gammie2001A}). Realistically, it is expected that upper layers of the gas cool down faster, resulting in a colder shadowed region able to produce a gradient pressure, promoting sharper spiral features for this mode.  For static shadows, we obtain similar results as in \cite{M+2016}. However,  in the prograde rotation case, a gas portion under the shadow moving at the same shadow's speed, will always be in the shade. This co-rotating position, can be obtained by equating the azimuthal velocity with the shadow speed $v_{\rm s} = (2\pi/T_{\rm s}) ~  r $. For a Keplerian disc, where $v_{\rm k} = \sqrt{G M / r}$, we obtain: $R_{\rm co} = \left( G M_\star (T_{\rm s} / 2\pi)^2  \right)^{1/3}$. In Figure~\ref{fig:resonance}, we plot $R_{\rm co}$ as a function of the shadow's period, and the azimuthal velocity along with the shadow speed for the different periods $T_{\rm s}$. The intersection between both curves gives the co-rotating location for each model (red dots). These points are in excellent agreement with the location of the local perturbations observed in the density fields shown in Figure \ref{fig:dens_press} (see also Table \ref{tab:simus}). We also found that the perturbation development  is independent of the stellar luminosity or the gas density.

The process by which prograde shadows promote planet-like spirals can be explained as follows:
If we set a reference frame at the co-rotating point $R_{\rm co}$ then the shadows are static. 
The disc rotates anti-clockwise in our models, therefore, in this frame, at $r < R_{\rm co}$ the azimuthal gas velocity is positive, 
and negative for $r > R_{\rm co}$.  The pressure gradients produced at the interfaces between 
shadowed and illuminated regions create a \textit{net force} acting in the negative azimuthal direction (clockwise)\footnote{cf. figure~3
from \cite{M+2016}.}. Therefore, at $r < R_{\rm co}$ (where $v_{\phi} > 0$) particles lose angular momentum,  
forcing them to move inwards. On the contrary, at $r > R_{\rm co}$ (where $v_{\rm \phi} < 0$) the gas gains 
angular momentum,  forcing outward accumulation. This results in the 
triggering of spiral in the density field, spreading inwards and outwards arms from the co-rotating location.

Interestingly, since the pressure field decreases with the radius, 
the force from the pressure gradient also decreases in the radial direction. Therefore, the loss of angular momentum 
in regions where $r < R_{\rm co}$ is larger than the gain of angular momentum at $r > R_{\rm co}$. 
This asymmetry explains why the inner arms are  more open (larger pitch angle) than the outer arms (smaller pitch angle) 
as noted in Table~\ref{tab:simus}. It is worth mentioning that the main morphological difference between 
spirals caused by planets and by moving shadows lies in the value of the pitch angle of the inner spirals.

The inner disc breaking likely happens because of the periodic gravitational perturbations of the inner stellar companion. It is worth highlighting that, once the shadows appear in the disc, one could use the spirals morphology to estimate the semi-major axis of the (potentially unseen) companion inside the central cavity (cf. Section~\ref{sec:precession}). Hence, this scenario constitutes a robust alternative to explain some enigmatic spirals observed in transition discs where embedded planets are still missing. In fact, this could very well explain why, despite the active search for companions, many spirals in transition discs remain uncorrelated with any planetary or stellar companion.

\section{Conclusions}
\label{sec:conclusions}

	We explored the hydrodynamical effect of rotating shadows on spiral formation in transition discs, extending the previous study by \cite{M+2016} to a broader range of disc configurations: static, retrograde and prograde shadows. The main results of this work can be summarized as follows:

\begin{itemize}
	\item The shadows movement affects the triggering and the morphology of the spirals: retrograde rotations do not produce any spirals in the disc, while static and prograde ones do. The shadowing effects could in principle be observed in scattered light (cf. Fig.~\ref{fig:obs}).
	\item For a broad range of values of $T_{\rm s}$, a prograde rotating shadow produce a spot in the density field, at exactly the co-rotating location (cf. Fig.~\ref{fig:resonance}), where the gas has the same orbital period as the rotating shadow.
	
	\item The morphology of the shadow-triggered spirals, notably resemble the planetary wakes caused by embedded planets in the disc, which are also characterized by two spiral arms. The pitch angle of the inner spiral is slightly lower in the planetary case, but the outer one is indistinguishable (see Table \ref{tab:simus}).
	\item Given that the shadow's period (or equivalently the precession period of the inclined inner disc) depends linearly on $T_{\rm b}$ (cf. Eq.~\ref{eq:Tp}), $T_{\rm s}$ can help to constrain the value of the binary semi-major axis for undetected companions. 
\end{itemize}
	
	Future observations, coupled to a better understanding of the underlying physics involved in planet formation, will allow to compare planetary and shadow-triggered spirals in protoplanetary discs in further detail. 
	
\section*{Acknowledgements}
We thank Jorge Cuadra for useful discussions throughout this project. M.M. and N.C. acknowledge financial support from Millenium Nucleus grant RC130007 (Chilean Ministry of Economy). N.C. acknowledges financial support provided by FONDECYT grant 3170680. We also thank the referee for constructive suggestions that have improved this letter.

\bibliographystyle{mnras}
\bibliography{movingshadows}

\end{document}